\documentclass[12pt]{article}
\input{epsf}
\usepackage{graphicx}
\def\AJ{{\it Astroph. J.} }

\def\CQG{{\it Class. Quantum Gravity} }

\def\GRG{{\it Gen. Relativity and Gravitation} }

\def\NP{{\it Nucl. Phys.} }
\def\PL{{\it Phys. Lett.} }
\def\PR{{\it Phys. Rev.} }

\def\frac#1#2{{\textstyle{{#1}\over {#2}}}}

\def\lsim{\mathrel{\rlap{\lower4pt\hbox{\hskip1pt$\sim$}}
    \raise1pt\hbox{$<$}}}
\def\gsim{\mathrel{\rlap{\lower4pt\hbox{\hskip1pt$\sim$}}
    \raise1pt\hbox{$>$}}}
\def\sqr#1#2{{\vcenter{\vbox{\hrule height.#2pt
         \hbox{\vrule width.#2pt height#1pt \kern#1pt
         \vrule width.#2pt}
         \hrule height.#2pt}}}}

\def\beq{\begin{equation}}
\def\eeq{\end{equation}}
\def\beqa{\begin{eqnarray}} 
\def\eeqa{\end{eqnarray}}

\def\laq{\raise 0.4 ex \hbox{$<$}\kern -0.8 em\lower 0.62 ex\hbox{$\sim$}}
\def\gaq{\raise 0.4 ex \hbox{$>$}\kern -0.7 em\lower 0.62 ex\hbox{$\sim$}}
\begin{document}
\titlepage

\begin{center}
{\bf Cosmology in Portugal: 
The First 20 Years}\footnote{Based on talk presented at the ``Astronomy and Astrophysics Research in Portugal: Its Impact 
and Future Perspectives'' meeting, 13 October 2006, Funda\c c\~ao Calouste Gulbenkian, Lisbon}
\vglue 1.5cm
{O. Bertolami\footnote{Also at Centro de F\'\i sica dos Plasmas, IST. E-mail address:orfeu@cosmos.ist.utl.pt}}

\bigskip
{\it Instituto Superior T\'ecnico,
Departamento de F\'\i sica,\\}
\medskip
{\it Av.\ Rovisco Pais 1, 1049-001 Lisboa, Portugal\\}

\vglue 1.5cm

\end{center}
\baselineskip=20pt

\centerline{\bf  Abstract}

\vglue 1cm

We reminisce on the first steps of the cosmology community in Portugal, which can be traced back to about 20 
years ago, and discuss its achievements and current specificities. We also reflect 
on the aspirations, hopes and challenges for the future.

\vglue 1cm
\noindent

\vfill
\newpage

\setcounter{equation}{0}
\setcounter{page}{2}

\baselineskip=20pt

\section{Reminiscences}

Cosmology is a relatively new science in Portugal. Its first steps  
can be traced back to about 20 years ago. This means that the development of cosmology as a research subject 
in Portugal has coincided with a particularly
blooming period, a phase in which cosmology has evolved from its more speculative origins into a  
subject with highly predictive theories and precision measurements. 

But, before we discuss the origins of cosmology in Portugal, we would better begin by paying a visit to the 
first discussions on relativity. Even though, in a passive way, Portugal has been related to the 
British expedition led by Arthur Eddington which was responsible for the measurement 
of the deflection of light in the 1919 eclipse. The measurements were carried out 
at Sobral, in the northeast region of Brazil, and at the Sundi 
area in Pr\'\i ncipe, then a Portuguese colony. That is to 
say, the confirmation of General Relativity occurred in two Portuguese speaking countries. 
In the case of the expedition to Brazil, there had been some 
participation of the Brazilian astronomers, and in what concerns the expedition 
to Pr\'\i ncipe, Eddington exchanged letters about logistics at the Sundi area
with Campos Rodrigues and Frederico Oom, respectively, director and vice-director of the 
Observat\'orio Astron\'omico de Lisboa \cite{Mota, Bertolami}.   

Actually, the first discussions about relativity in Portugal were of philosophical nature, and are due to 
Leonardo Coimbra (1883-1935) from the Faculdade de Letras da Universidade de Lisboa in 1912, on a 
dissertation about creationism. From then on, in the 1920s, the discussion was conducted predominantly by mathematicians, 
followed by a period of philosophical arguing and polemics about the interpretation of specific issues within 
relativity in the 1930s. The first lectures on Special and General Relativity in Portugal took place at the 
mathematics major at the Faculdade de Ci\^encias da Universidade de Lisboa in the year 1922-1923, delivered by 
Ant\'onio dos Santos Lucas (1866-1939). The first research papers on Relativity, actually on differential geometry 
inspired by General Relativity and Unified Field Theory, are due to 
Aureliano Mira Fernandes (1884-1958) from Instituto Superior T\'ecnico and appeared from 1928 till 1937 
(at a rate of at least one per year)
in the prestigious Italian journal {\it Rendiconti da Academia dei Lincei}. This is a most remarkable and 
singular endeavour in the scenario of the Portuguese academic life at the time. Another important name at the 
time was Rui Lu\'\i s Gomes, a disciple of Mira Fernandes who published in the {\it Rendiconti da Academia dei Lincei} a 
paper about the Special Theory of Relativity and created at the Faculdade de Ci\^encias da Universidade do Porto, a theoretical 
physics seminar, which was the first organized attempt to set up, in a concerted fashion, theoretical physics in Portugal. 
In 1930, in order to fulfill the requirement of the examination 
for professorship at the University of Coimbra\footnote{The University of Coimbra is the oldest in the country. It was established 
through a decree by King Dom Dinis and confirmed by the pope Nicolau IV in 9 of August of 1290.}, 
Manuel dos Reis (1900 - 1992), an astronomer and mathematician, 
wrote an influential monograph, {\it O Problema da Gravita\c c\~ao Universal}, where gravity and relativity 
were discussed in an historical perspective \cite{Fitas}. During the academic year of 1930-1931, M\'ario Silva (1901-1977), 
who obtained a Ph.D. with Mme. Curie and was back to Coimbra in that year, 
declared his intention ``to launch a discussion within the tiny Coimbra scientific community ... on some new doctrines 
such as the quantum and relativity theories'' and which involved Manuel dos Reis.
These discussions would lead to the first public confrontation between anti and pro-relativists in the country. 
The polemics took place in the journal {\it Seara Nova} (New Grain Field) and involved M\'ario Silva and 
Admiral Gago Coutinho\footnote{Admiral Gago Coutinho (1869-1959) and the airman Artur de Sacadura Cabral (1880-1924), 
were the first men to cross the South Atlantic in a non-stop flight from Lisbon to Rio de Janeiro in 1922.}, 
who had heard Einstein lectures in Brazil in 1925 and attended the lectures 
of Paul Langevin in Lisbon in late 1929, but could not accept or understand 
some of the kinematic implications of Special Relativity. 
Other pro-relativistic texts appeared in the cultural journal {\it O Diabo} (The Devil) by Abel Salazar\footnote{Abel 
Salazar (1889-1946) was a physician, scientist, author and artist who lived in Porto and was an influential 
figure in the Portuguese cultural life from 1920s onward.} \cite{Fitas}. Influential textbooks 
on electromagnetism and special relativity appeared in the second half of the 1940s by 
M\'ario Silva \cite{Silva} from Coimbra and  Ant\'onio da Silveira (1904-1985) \cite{Silveira} 
from Instituto Superior T\'ecnico. From the 1940 to the 1960s, the central figure in relativity and cosmology in Portugal was 
Ant\'onio Gi\~ao from the Faculdade de Ci\^encias da Universidade de Lisboa and Instituto Gulbenkian de 
Ci\^encias. Gi\~ao has exchanged letters with 
Einstein, published papers on relativity and cosmology in the Physical Review \cite{Giao1,Giao2} and organized a NATO 
cosmology meeting in 1963, which had among the participants, Pascual Jordan, Yves Thiry, Hermann Bondi and others \cite{Giao3}.  
To this scarce, but regular interest in relativity, 
followed a quiet period till the {\it Revolu\c c\~ao dos Cravos} in 25th of April 1974, when a military coup restored democracy in 
Portugal after almost 50 years of dictatorship. Since then, the universities and the academic life have undergone profound
transformations. The university population has, since then, grown about eight-fold and academic standards have 
improved considerably, despite the still rather limited resources\footnote{Even at present, the fraction of the GNP allocated 
to research and development does not exceed about $0.8 \%$, one of the lowest of the European Union.}. 
Interest in relativity has awakened again in the mid 1970s and a course 
on General Relativity was then delivered by the French physicist Maurice Bazin 
based on the textbook of which he is a co-author \cite{Bazin} 
at the Faculdade de Ci\^encias da Universidade de Lisboa. Subsequently, courses were delivered in the same faculty 
by Ant\'onio Brotas from Instituto Superior T\'ecnico in late 1970s \cite{Brotas}. 
The first course on Relativity, which included 
also Cosmology was delivered by  Paulo Crawford do Nascimento in autumn/winter semester of 1983 in the physics course 
in the Faculdade de Ci\^encias da Universidade de Lisboa. 
In Instituto Superior T\'ecnico, the first Relativity 
and Cosmology course was held by this author to students of the Physics and Mathematics 
Departments in the spring/summer semester of 1993. 
About the same time, cosmology lectures started being held by  Paulo Gali Macedo at the Universidade do Porto.      
Since then, lectures on General Relativity and Cosmology have been regularly held in these three universities, both at undergradute 
and graduate levels.

The first papers on cosmology were written in Portugal about 20 years ago. The first ones were solo flights 
by Alfredo Barbosa Henriques concerning higher dimensional cosmological models \cite{Alfredo1}, 
by Estelita Vaz on Bianchi-type spaces \cite{EVaz} and a more mathematical work by Maria Helena Bugalho, 
Amaro Rica da Silva and Jos\'e Sousa Ramos on chaos on a Bianchi-IX cosmological model \cite{Bugalho1}. Three papers appeared 
in the following year, another one by Alfredo Barbosa Henriques in collaboration with Gordon Moorhouse \cite{Alfredo2}, 
a paper by Paulo Crawford Nascimento in collaboration with Rubem Mondaini on anisotropic universes and 
inflation \cite{PCNascimento} and a paper by Maria Helena Bugalho on cosmologies with a non-Abelian two-parameter isometry group 
\cite{Bugalho2}. In this period, research activities were carried 
out at the Instituto de F\'\i sica e Matem\'atica, now Complexo Multidisciplinar da Universidade de Lisboa at Avenida Gama Pinto, 2, 
a multidisciplinary and multi-university research institute which encompassed 
researchers in atomic and nuclear physics, condensed matter physics, field theory, particle physics and mathematics, from the three 
Lisbon universities: Universidade de Lisboa, Universidade T\'ecnica de Lisboa and Universidade Nova de Lisboa. The institute 
existed since the 1960s and was founded by Ant\'onio da Silveira, funded initially by the Instituto da Alta Cultura, a 
government financing institution of cultural and scientific activities, and 
later on by the Instituto Nacional de Investiga\c c\~ao Cient\'\i fica (INIC), a national research foundation which existed till 
1992. By the late 1980s and early 1990s, there was a convergence of people coming from abroad with a wide range of interests 
including cosmology. The mathematical physicist Roger Picken had been in the institute since 1987. 
This author did arrive from a post-doctoral experience in Heidelberg 
to work at the Instituto de F\'\i sica e Matem\'atica in October 1989 
with a post-doctoral fellowship from Grupo Te\'orico de Altas Energias; Jos\'e Mour\~ao arrived in 1990 after completing his Ph. D. 
studies in Moscow and so did Maria da Concei\c c\~ao Bento, who came from a post-doctoral period at London. The four of us 
were in the early 1990s hired as lecturers at Instituto Superior T\'ecnico at the Mathematics and Physics Departments.
Together with two young Ph. D. students, Paulo Moniz and Paulo S\'a and a M. Sc. student, Lu\'\i s Mendes, 
it was possible to organize a weekly 
seminar in cosmology, to build a preprint library and to create an environment where various problems of cosmology could be 
tackled and discussed. These included, for instance, solutions of the Einstein-Yang-Mills system (classical \cite{MonizM}, 
wormhole-like \cite{BMPV} and quantum cosmological \cite{BMo}), Kantowski-Sachs cosmological models 
\cite{ABHM}, Bianchi IX models \cite{ABHMS}, symmetry breaking in curved spacetime \cite{MCB} and string cosmology 
\cite{BBS1}. By 1993, Jos\'e Pedro Mimoso \cite{Mimoso} was back to Faculdade de Ci\^encias of the Universidade de Lisboa after 
completing his Ph. D. in Sussex.

The other major center of activity in cosmology in Portugal is Porto where the first papers to appear were
on bounds for the Jordan-Thiry scalar field coupling constant by Paulo Gali Macedo \cite{Macedo}, 
on the Sunyaev-Zeldovich effect and its cosmological implications \cite{DBB} in 1996 by Domingos Barbosa and co-authors, 
on cosmic strings \cite{PA} in 1997 by Pedro Avelino and co-authors 
and, on galaxy cluster abundance and cosmological parameters \cite{PV} in 1999 by  Pedro Viana and Andrew Liddle. 
More recently, Carlos Herdeiro and Miguel Costa and Carlos Martins 
have joined the group and enriched the list of research subjects 
to include G\"odel type universes \cite{CH02}, string cosmology \cite{MC03} and  
cosmic strings \cite{AM} to mention just a few. 

The other groups, although smaller, are no less relevant. The group at Universidade do Minho, got established in the 
early 1980s after Estelita Vaz came back from her Ph. D. in Britain; Jos\'e Castanheira da
Costa joined her a few years later and went to the Universidade da Madeira in the 1990s. 
The group at the Universidade do Algarve was established in the early 
1990s by Cenalo Vaz and Robertus Potting \cite{KRP}, the former now gone to the US, 
and Paulo S\'a joined the group in 1995 \cite{PS}. The group at 
the Universidade da Beira Interior was established by Paulo Moniz in the late 1990s \cite{LM} 
and Jos\'e Velhinho joined 
him in 2000 \cite{Velhinho}.

\section{Current Trends}

Currently, research in cosmology in Portugal is carried out in 6  
out of the 14 state Universities\footnote{The public sector also comprises about 20 polytechnic institutions.}. 
They are the following, where we indicate 
within parenthesis the name of the researchers:

\vglue 0.3cm

\noindent
Universidade do Algarve at Faro (Paulo S\'a and Robertus Potting) 

\vglue 0.3cm

\noindent
Universidade da Beira Interior at Covilh\~a (Paulo Moniz and Jos\'e Velhinho)

\vglue 0.3cm

\noindent 
Universidade de \'Evora (Il\'\i dio Lopes)

\vglue 0.3cm

\noindent
Faculdade de Ci\^encias da Universidade de Lisboa (Luis Bento, Paulo Crawford, Tom Girard and Jos\'e Pedro Mimoso and Ana Nunes)

\vglue 0.3cm

\noindent 
Instituto Superior T\'ecnico of the Universidade T\'ecnica de Lisboa (Maria da Concei\c c\~ao Bento, Orfeu Bertolami, 
Alfredo Barbosa Henriques and Ana Mour\~ao) 

\vglue 0.3cm

\noindent
Universidade do Minho at Braga and Guimar\~aes (Filipe Mena, Piedade Ramos and Estelita Vaz)

\vglue 0.3cm

\noindent
Faculdade de Ci\^encias da Universidade do Porto (Pedro Avelino, Miguel Costa, Paulo Carvalho, Carlos Herdeiro, 
Catarina Lobo, Paulo Gali Macedo, F\'atima Mota, Caroline Santos and Pedro Viana)

\vglue 0.2cm
Quite often, the research is developed in Physics Departments, but also at Applied Math Departments in the 
case of the Universidade do Porto and Universidade do Minho. However, formally, and this is a well marked feature of 
science activities in Portugal, research activities take place in the context of research centers which are related with the 
universities, but have some administrative independence and some freedom to guide research interests 
and their development. Cosmology is a research topic in 
8 centers. Their spread within the above mentioned universities is as follows: 

\vglue 0.3cm

\noindent
Faculdade de Ci\^encias da Universidade de Lisboa:

\noindent
Centro de Astronomia e Astrof\'\i sica da Universidade de Lisboa (CAAUL)

\noindent
Centro de F\'\i sica Nuclear  (CFN) 

\noindent 
Centro de F\'\i sica Te\'orica e Computacional (CFTC).

\vglue 0.3cm

\noindent
Instituto Superior T\'ecnico (IST):

\noindent
Centro Multidisciplinar de Astrof\'\i sica (CENTRA)

\noindent
Centro de F\'\i sica dos Plasmas (CFPl) 

\noindent
Centro de F\'\i sica Te\'orica e Part\'\i culas (CFTP).

\vglue 0.3cm

\noindent
Faculdade de Ci\^encias da Universidade do Porto:

\noindent
Centro de Astrof\'\i sica da Universidade de Porto (CAUP) 

\noindent
Centro de F\'\i sica do Porto (CFP).

\vglue 0.2cm

Besides the above mentioned researchers, an important feature of the Portuguese 
scientific community working in cosmology is that it is relatively young and that 
a great deal of activity is carried out by young researchers, post-docs and Ph.D. students.
There are about 18 post-docs and their distribution around the country is as follows: 
1 in Braga-Guimar\~aes, 1 in Faro, 12 in Lisbon, 4 in Porto and there are 6 Portuguese post-docs 
abroad (indicated with *), some 
willing to come back within a period of 1 to 3 years. Let me name them:

\noindent
Tiago Barreiro, Domingos Barbosa, Jarle Brinchmann, Simone Calogero,
Carla Carvalho, Patricia Castro, Tiago Charters, Morgan Le Delliou, 
Vikram Duvvuri, S\'ebastien Fabbro, Pedro C. Ferreira,
Francisco Lobo, Carlos Martins, Luis Mendes*, David Mota*, Rui Neves, 
Nelson Nunes*, Jorge P\'aramos, 
M\'ario Santos, Nuno Santos*, Ant\'onio Silva, Pedro Silva, 
Ricardo Schiappa*, Luis Teodoro*. 

\vglue 0.2cm

There are also about 14 Ph. D. students: 1 in Braga-Guimar\~aes, 1 in Faro, 5 in Lisbon and 7 in Porto. Let me name them:

\noindent
Vladan Arsenivejic, Luis Be\c ca, Catarina Bastos, Irene Brito, Lu\'\i s Costa, 
Rosa Doran, Jo\~ao Fernandes, Josinaldo Menezes, 
Joana Oliveira, Carmen Rebelo, 
Ismael Tereno, Brigitte Tom\'e, Paulo Torres, Pedro Vieira.

\vglue 0.2cm

Currently, the research in cosmology in Portugal covers the major areas of interest, which includes 
quantum cosmology, strings and brane cosmology, inflation, cosmic microwave background radiation, cosmic strings, 
cosmological phase transitions and cosmology in the laboratory, 
dark matter, dark energy, variation of fundamental couplings, spacetime symmetries and topology 
and observational cosmology.

In many of these subjects, papers written in Portugal have reached a position of visibility, having 
more than 50 and 100 citations according to the QSPIRES database. Among those, one could mention, for 
instance, papers on baryogenesis \cite{BCKP}, cosmological and astrophysical aspects of the 
of Lorentz symmetry breaking \cite{BC,BMota}, dark energy \cite{Bento1,BM, Bento2, Bento3,BSSS}, 
inflation \cite{Bento6} and the variation of fundamental constants \cite{AMRV1,AMRV2,BLPR}. Papers on Lorentz symmetry breaking 
and gravity \cite{KP} and on dark energy \cite{Bento4} have been awarded 3rd Prizes by 
the Gravity Research Foundation in U.S., in 1999 and 2005. A paper on dark matter has been awarded the 
Prize Uni\~ao Latina de Ci\^encia in 2001 \cite{Bento5}. 

It is quite exciting that this quality recognition can also be found in papers on observational cosmology and data analysis, 
such as, for instance, 
on the Sunyaev-Zeldovich effect and its implications for cosmology \cite{BD}, on  
constraints on the matter power spectrum normalization using SDSS/RASS and reflex cluster survey \cite{LV02} and on 
supernova survey and cosmological parameters \cite{Many}. It is also quite remarkable the Portuguese involvement on the team that has 
identified the oldest cluster ever observed, with $z=1.45$, in the context of the XMM cluster survey \cite{LV06}.

\vglue 0.2cm

These papers indicate that, recently, there has been a serious involvement of the Portuguese cosmology community 
on observational activities. The most salient are the following:

\vglue 0.3cm

\noindent
Galactic Emission Mapping (Building a radio telescope for studying the foreground of the galactic synchrotron emission) (CENTRA)

\vglue 0.3cm

\noindent
Planck Surveyor (CENTRA-Working Group 7, CAUP-Working Group 5)

\vglue 0.3cm

\noindent
Square Kilometer Array (CENTRA)

\vglue 0.3cm

\noindent
Supernova Cosmology Project and Supernova Legacy Survey (CENTRA)

\vglue 0.3cm

\noindent
XMM-Newton Cluster Survey (CAUP)

\vglue 0.3cm

\noindent
Pioneer Anomaly and LATOR (Laser Astrometric Test of Relativity) Science Teams (CFPl)

\vglue 0.3cm

We can also mention the involvement of the community on recent publications aiming to raise the public 
awareness on cosmology and, in particular, on the work developed in Portugal: 

\vglue 0.3cm

\noindent
{\it Descobrir o Universo}, Teresa Lago {\it et al.} (Ed. Gradiva, April 2006).

\vglue 0.3cm

\noindent
{\it O Livro das Escolhas C\'osmicas}, Orfeu Bertolami (Ed. Gradiva, January 2006).

\vglue 0.2cm

\noindent
To these books, we can add the chapter on cosmology written by Jos\'e Pedro Mimoso which appeared in the book 
{\it O C\'odigo Secreto} (Ed. Gradiva) by Margarida Telo da Gama {\it et al.} in 2005.

\vglue 0.2cm

Portugal has been also the stage of many international conferences on cosmology. Indeed, since the 1990s, conferences have 
been taken place in Portugal on a regular basis. The most attended include: 

\vglue 0.2cm

\noindent
XII Autumn Lisbon School: The Physical Universe, Lisbon, October 1990. 

\noindent
Orgs. J. Barrow, A. B. Henriques, M. T. V. Lago and M.S. Longair \cite{BHLL}

\vglue 0.3cm

\noindent
Iberian Meeting on Gravity, \'Evora, September 1992.

\noindent
Orgs. M. C. Bento, O. Bertolami, A. B. Henriques, J. Mour\~ao and R. Picken \cite{BBMP}

\vglue 0.3cm

\noindent
Electroweak Physics and the Early Universe, Sintra, March 1994.

\noindent
Orgs. Jorge C. Rom\~ao and Filipe Freire \cite{RF}

\vglue 0.3cm

\noindent
Non-Sleeping Universe Conference, Porto, November 1997.

\noindent
Orgs. M. T. V. Lago and A. Blanchard \cite{LB}

\vglue 0.3cm

\noindent
Cosmology 2000, Lisbon, July 2000.

\noindent
Orgs. M. C. Bento, O.Bertolami, A. B. Henriques and L. Teodoro 

\noindent
$(http://alfa.ist.utl.pt/~bento/cosmo2000/cosmo.html)$

\vglue 0.3cm

\noindent
JENAN 2002 The Unsolved Universe: Challenges for the Future, Porto, September 2002. 

\noindent 
$(http://www.sp-astronomia.pt/jenam2002/)$

\vglue 0.3cm

\noindent
The Quest for Cosmological Scalar Fields, Porto, July 2004. 

\noindent
$(http://www.fc.up.pt/pessoas/luis.beca/)$

\vglue 0.3cm

\noindent
School on Superstrings: New Trends in String Theory, Lisbon and Porto (1998, 2001, 2004).

\noindent
Orgs. M. C. Bento, O. Bertolami, M. Costa, C. Herdeiro, J. Mour\~ao, J. Pimentel and R. Schiappa

\noindent
$(http://www.math.ist.utl.pt/~strings/MTST2/)$

\vglue 0.3cm

\noindent
New Worlds in Astroparticle Physics, Faro (1996, 1998, 2000, 2002, 2005).   

\noindent
Orgs. J. Dias de Deus, A. Mour\~ao, M. Pimenta, R. Potting and P. S\'a

\noindent
$(http://www.ualg.pt/fct/fisica/centra/astroparticle_2005/a2005.html)$

\vglue 0.3cm

\section{The Future: Aspirations and Challenges}

It is clear that cosmology has undergone an impressive development in Portugal in the first two decades of its history. 
However, despite this growth, there are serious concerns about the will of the institutions 
to keep up this positive trend in the future. The reasons are manifold, the most noticeable being:    

\vglue 0.3cm

\noindent
i) The university and the national laboratory system are still too opaque and have shown no clear signs 
that they are able and willing to absorb the young generation via fresh job positions and to create 
intellectually stimulating research spaces. In our opinion, the creation of stable positions for the young generation 
of researchers should be a top priority. 
Furthermore, several structural changes must be 
promoted so that careers are regulated by criteria 
of excellence and not by ``Monte Carlo'' methods ... That is to say, a radical change is urgently 
needed in the criteria of access and promotion for the university and research positions. 
Emphasis on creativity, innovation and mobility 
within institutions must be implemented as soon as possible so as to avoid a brain drain problem.

\vglue 0.3cm

\noindent
ii) It is unclear whether the cosmology community will be able to rely on a regular and continuous support 
from the existing institutions (Minist\'erio da Ci\^encia, Tecnologia e Ensino Superior, Funda\c c\~ao para a Ci\^encia e a 
Tecnologia and so on) given budgetary difficulties arising from restrictions by the European Monetary Fund, economic 
performance and so on. Furthermore, financial support is often incipient and too conditioned by changes of
political nature.    

\vglue 0.3cm

We think that it is uncontroversial to state that it would be beneficial for 
the community to count on with a national laboratory devoted to Astronomy, Astrophysics and 
Cosmology given that there are no 
astronomy departments in the country. Of course, an institution of this nature should concentrate various competences such as, 
for instance, the ones associated with data storage, computational facilities and hardware support. 
A national laboratory would also help to strengthen the spirit of the community
as well as the ties with the existing financing institutions and could open up new forms of collaboration 
with private institutions and individuals. Finally, we feel that it is important to stress that, despite all the 
difficulties, the cosmology community in Portugal can 
realistically aim to establish its position in the league of the top 5 strongest in Europe. The assessment of the first 20 
years of development indicates that this is quite within reach, however this goal can only be 
achieved if we can ensure, through our concerted action, the long term success of the portuguese 
cosmology community and if conditions are created for the settling and 
blooming 
within our institutions of the young generation of researchers. It would be a 
waste if the country could not make proper use of the existing talents.


\vfill

\centerline{\bf {Acknowledgments}}

\vskip 1 cm

\noindent
I would like to express my gratitude to Miguel de Avillez, Andr\'e Moutinho and Ant\'onio Pedrosa, members 
of the present direction of the Sociedade Portuguesa de Astronomia (SPA), 
for the invitation to deliver this talk and  
to Pedro Avelino, Domingos Barbosa, Paulo Crawford, Jorge Dias de Deus, Alfredo Barbosa Henriques, 
Jos\'e Pedro Mimoso, Ana Mour\~ao and 
Ant\'onio Silva for the information about their centers and their research activities. 
I am also indebted to Sonia Anton and 
Jo\~ao Fernandes, members of the former direction of SPA, for having laid out the first stone of the White Book of   
Astronomy and Astrophysics in Portugal. Finally, I would like to 
acknowledge the partial support of Funda\c c\~ao para a 
Ci\^encia e a Tecnologia under the grant POCI/FIS/56093/2004.

\vfill
\newpage


\end{document}